\RequirePackage{fix-cm}
\documentclass[smallextended]{svjour3}
\makeatletter\if@twocolumn\PassOptionsToPackage{switch}{lineno}\else\fi\makeatother

  \makeatletter
  \def\fig@textbf{\textbf}
   \AtBeginDocument{\renewcommand\floatc@plain[2]{\setbox\@tempboxa\hbox{\textbf{\footnotesize#1.}\footnotesize\hskip.5em#2}%
    \ifdim\wd\@tempboxa>\hsize {\fig@textbf{\footnotesize#1.}}\footnotesize\hskip.5em#2\par
        \else\hbox to\hsize{\hfil\box\@tempboxa\hfil}\fi}}
    \makeatother
  
\usepackage{amsfonts,amssymb,amsbsy,tabulary,amsmath}
\smartqed  
\usepackage{graphicx}
\usepackage{url,multirow,morefloats,floatflt,cancel,tfrupee}
\makeatletter

\AtBeginDocument{\@ifpackageloaded{textcomp}{}{\usepackage{textcomp}}}
\makeatother
\usepackage{colortbl}
\usepackage{xcolor}
\usepackage{pifont}
\usepackage[nointegrals]{wasysym}
\makeatletter

\def\mcWidth#1{\csname TY@F#1\endcsname+\tabcolsep}

\def\cAlignHack{\rightskip\@flushglue\leftskip\@flushglue\parindent\z@\parfillskip\z@skip}
\def\rAlignHack{\rightskip\z@skip\leftskip\@flushglue \parindent\z@\parfillskip\z@skip}

\@ifundefined{etal}{}{}

\usepackage{ifxetex}
\ifxetex\else\if@twocolumn\@ifpackageloaded{stfloats}{}{\usepackage{dblfloatfix}}\fi\fi

\AtBeginDocument{
\expandafter\ifx\csname eqalign\endcsname\relax
\def\eqalign#1{\null\vcenter{\def\\{\cr}\openup\jot\m@th
  \ialign{\strut$\displaystyle{##}$\hfil&$\displaystyle{{}##}$\hfil
      \crcr#1\crcr}}\,}
\fi
}

\AtBeginDocument{%
  \@ifpackageloaded{endfloat}%
   {\renewcommand\efloat@iwrite[1]{\immediate\expandafter\protected@write\csname efloat@post#1\endcsname{}}}{\newif\ifefloat@tables}%
}%

\def\BreakURLText#1{\@tfor\brk@tempa:=#1\do{\brk@tempa\hskip0pt}}
\let\lt=<
\let\gt=>
\def\processVert{\ifmmode|\else\textbar\fi}

\@ifundefined{subparagraph}{
\def\subparagraph{\@startsection{paragraph}{5}{2\parindent}{0ex plus 0.1ex minus 0.1ex}%
{0ex}{\normalfont\small\itshape}}%
}{}

\newcommand\role[1]{\unskip}
\newcommand\aucollab[1]{\unskip}
  
\@ifundefined{tsGraphicsScaleX}{\gdef\tsGraphicsScaleX{1}}{}
\@ifundefined{tsGraphicsScaleY}{\gdef\tsGraphicsScaleY{.9}}{}
\def\checkGraphicsWidth{\ifdim\Gin@nat@width>\linewidth
	\tsGraphicsScaleX\linewidth\else\Gin@nat@width\fi}

\def\checkGraphicsHeight{\ifdim\Gin@nat@height>.9\textheight
	\tsGraphicsScaleY\textheight\else\Gin@nat@height\fi}

\def\fixFloatSize#1{}
\let\ts@includegraphics\includegraphics

\def\inlinegraphic[#1]#2{{\edef\@tempa{#1}\edef\baseline@shift{\ifx\@tempa\@empty0\else#1\fi}\edef\tempZ{\the\numexpr(\numexpr(\baseline@shift*\f@size/100))}\protect\raisebox{\tempZ pt}{\ts@includegraphics{#2}}}}

\AtBeginDocument{\def\includegraphics{\@ifnextchar[{\ts@includegraphics}{\ts@includegraphics[width=\checkGraphicsWidth,height=\checkGraphicsHeight,keepaspectratio]}}}

\DeclareMathAlphabet{\mathpzc}{OT1}{pzc}{m}{it}

\def\URL#1#2{\@ifundefined{href}{#2}{\href{#1}{#2}}}

\def\UrlOrds{\do\*\do\-\do\~\do\'\do\"\do\-}%
\g@addto@macro{\UrlBreaks}{\UrlOrds}

\edef\fntEncoding{\f@encoding}

\makeatother

\newif\ifmultipleabstract\multipleabstractfalse%
\usepackage[numbers,sort&compress]{natbib}

\usepackage[T1]{fontenc}

\makeatletter	

\AtBeginDocument{%
  \@ifpackageloaded{endfloat}%
   {
\renewcommand*\efloat@process[2]{%
  \ef@ifct{#1}{%
    \expandafter\immediate\expandafter\closeout\csname efloat@post#1\endcsname
    \ef@setct{#1}{0}%
    \clearpage                                                         
        
    \efloat@ifflag{#2list}{
      {\normalsize\efloat@listof{#2}}
    }{}%

    \efloat@ifflag{#2head}{%
      \section*{\@nameuse{#2section}}
      \suppressfloats[t]
    }{}

    \markboth                                                          
      {\expandafter\uppercase\expandafter{\csname #2section\endcsname}}
      {\expandafter\uppercase\expandafter{\csname #2section\endcsname}}

    \def\efloat@type{#2}%
    \processdelayedfloat@hook
    \@nameuse{process#2s@hook}%
    \clearpage
    \@input{\jobname.#1}%
  }{}}
   
   }{}%
}%

  \makeatother

\usepackage{float}

\begin{document}

\title{Automated Segmentation of Vertebrae on Lateral Chest Radiography Using Deep Learning}\author{\raggedright Sanket~Badhe \and Varun~Singh \and Joy~Li \and Paras~Lakhani}
\institute{Sanket~Badhe \at Computer Science Department\unskip, Rutgers University\\\email{sanket.badhe@rutgers.edu} \and Varun~Singh \at Department of Radiology\unskip, Thomas Jefferson University 
    \\\email{vxs039@jefferson.edu} \and Joy~Li \at \\\email{joy.li@jefferson.edu} \and Paras~Lakhani \at Department of Radiology\unskip, Thomas Jefferson University 
    \\\email{M.D.}}\titlerunning{{Automated Segmentation of Vertebrae on Lateral Chest Radiography Using Deep Learning}}

\authorrunning{Sanket~Badhe et al.}
        \date{Received: date / Accepted: date}

      \maketitle

\begin{abstract}
The purpose of this study is to develop an automated algorithm for thoracic vertebral segmentation on chest radiography using deep learning. 124 de-identified lateral chest radiographs on unique patients were obtained.  Segmentations of visible vertebrae were manually performed by a medical student and verified by a board-certified radiologist.  74 images were used for training, 10 for validation, and 40 were held out for testing. A U-Net deep convolutional neural network was employed for segmentation, using the sum of dice coefficient and binary cross-entropy as the loss function. On the test set, the algorithm demonstrated an average dice coefficient value of 90.5 and an average intersection-over-union (IoU) of 81.75. Deep learning demonstrates promise in the segmentation of vertebrae on lateral chest radiography. \mbox{~}
\end{abstract}
    
\section{Purpose }
The purpose of this study is to develop an automated, validated algorithm for vertebral body segmentation on lateral chest radiographs using deep learning. Successful, automated segmentation of vertebral bodies could lead to the detection of spinal fractures and the precise quantification of vertebral body heights. Automated vertebral body height measurements appropriately applied to a large, diverse dataset could stratify mean values and standard deviations of vertebral heights by patient age, height, sex, and other clinical parameters.
    
\section{Introduction}
In the United States, an estimated 750,000 new spinal compression fractures occur every year. On routine lateral chest radiography, vertebral fractures and compression deformities may be under-diagnosed and sometimes difficult to appreciate. The progression of undetected fractures places patients at an increased risk for complications associated with significant morbidity and mortality \unskip~\cite{406596:9006088}  An automated, validated solution to the detection of vertebral body compression fractures on lateral chest radiographs may facilitate  early diagnosis and allow for timely intervention to unburden patients of preventable compression fracture sequela. 

Prior automated solutions for vertebral segmentation using traditional computer aided detection (CAD) approaches, demonstrated modest results, likely due to the sheer heterogeneity of patient anatomical variation. A study performed by Mysling et al. outperformed previous vertebral segmentation attempts but demonstrated a 52\% error rate in the  segmentation of fractured vertebrae and a 10\% overall vertebral segmentation error rate. The efficacy of this approach was analyzed incompletely, but segmentation error was defined in this study as {\textgreater}2 millimeters point-to-countor distance, a length metric between the ground truth manual segmentation and the algorithm-based segmentation.  Wong et al. proposed a live fluoroscopic vertebral segmentation strategy, but demonstrated no performance data to assess the technique's real-world performance.\unskip~\cite{406596:9112445}

Deep learning techniques have demonstrated  success in pixel-level labeling tasks using semantic segmentation, which is the partition of an image into unique parts or objects, such as identifying all cars or people within an image. \unskip~\cite{406596:12625730} In medical images, this has been applied to  the segmentation of pancreatic tumors on CT \unskip~\cite{406596:9006089} , brain tumors on MRI \unskip~\cite{406596:9006176}, and stroke lesions on MRI for example. \unskip~\cite{406596:9006177} Some deep learning methods for semantic segmentation include fully convolutional neural networks (FCN) \unskip~\cite{406596:8996620}  convolutional autoencoders such as the U-Net \unskip~\cite{406596:8996836}  and DeepLab \unskip~\cite{406596:8999026} .  As opposed to semantic segmentation, there also exist solutions using deep learning for instance segmentation, where each object instance is identified within an image (e.g. car 1, car 2, car 3, etc...).  For example, with regard to vertebrae, an instance segmentation solution could identify each vertebral body such as T1, T2, T3 and color-code them for example.  Some popular instance segmentation solution includes Mask-RCNN.\unskip~\cite{406596:12514912}

In this study, we choose to employ semantic segmentation using a standard U-Net\unskip~\cite{406596:8996836} , as it has shown to be relatively accurate with regard to segmentation in medical imaging, including segmentation of brain tumors, kidneys and pulmonary nodules.\unskip~\cite{406596:9004564,406596:9004648,406596:9004690}    The U-Net network architecture is structured into an encoder and a decoder. The encoder follows the classic architecture of the convolutional neural network, with convolutional blocks each followed by a rectified linear unit (ReLU) and a max polling operation to encode image features at different levels of the network. The decoder up-samples the feature map with subsequent up-convolutions and concatenations with the corresponding encoder blocks. This network style architecture helps to better localize and extract image features and assembles a more precise output based on encoder information.

The model is first created using the training dataset by comparing the training  data with expected output to establish optimal weights  with back-propagation rules. Validation data is then used to establish the optimal number of hidden units to verify a stopping point for the back-propagation algorithm of the trained model essential for model selection. The test dataset is utilized to establish the accuracy of the model from the fully trained final model weights. The test and validation datasets are categorized independently to ensure accuracy as the final model is biased toward the validation data used to make final model selection.

\bgroup
\fixFloatSize{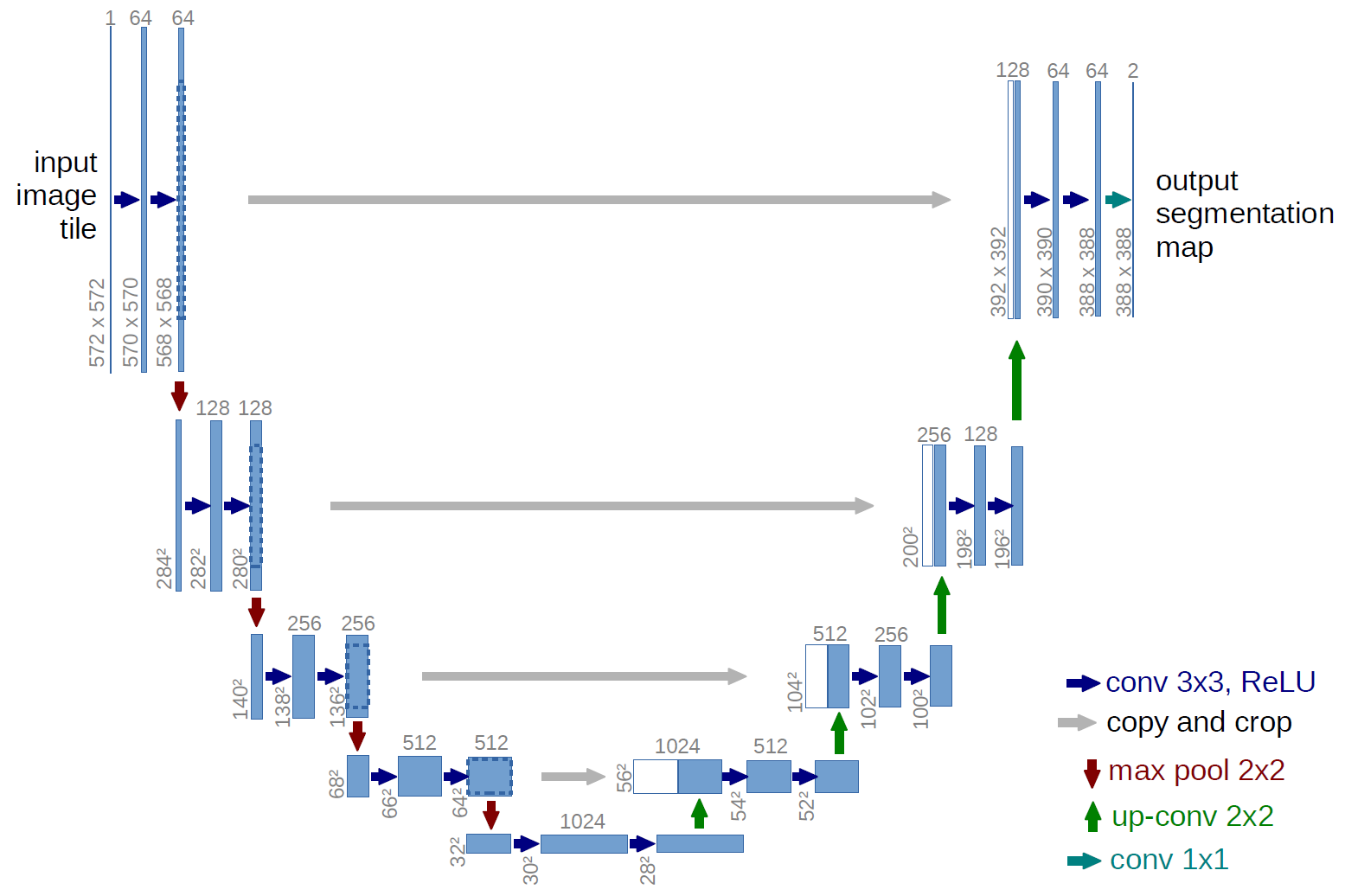}
\begin{figure*}[!htbp]
\centering \makeatletter\IfFileExists{images/4f74987c-0a52-400f-a026-b3b1ad7b4373-u2.png}{\includegraphics{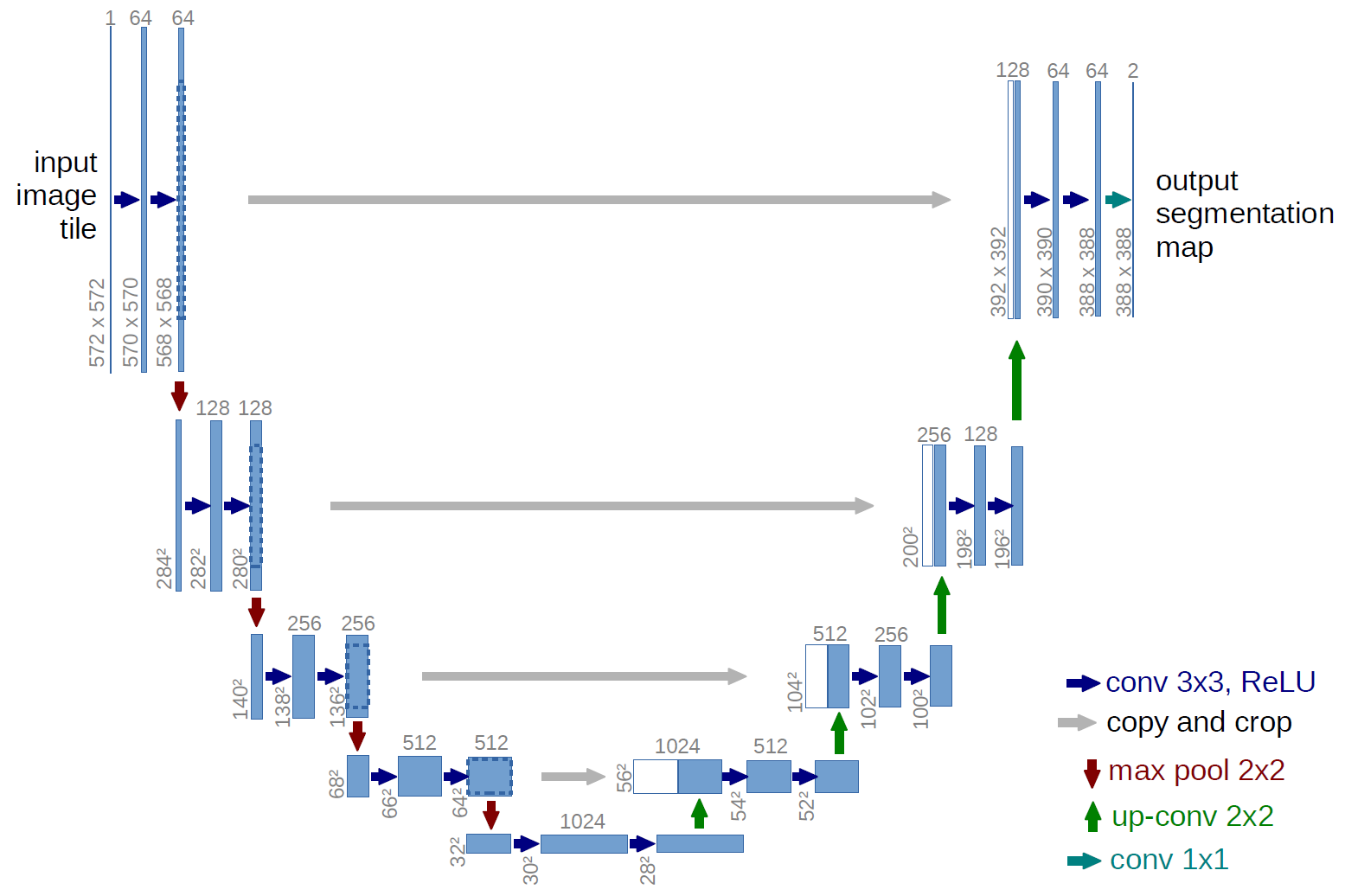}}{}
\makeatother 
\caption{{\textit{U-Net Architecture}}}
\label{f-e5f3}
\end{figure*}
\egroup
In Figure~\ref{f-e5f3}, each blue box corresponds to a multi-channel feature map. The number of channels is denoted on top of the box. The x-y-size is provided at the lower left edge of the box. White boxes represent copied feature maps. The arrows denote the different operations. 

As vertebrae typically represent less than 20\% of the total area in each lateral radiograph, vertebral segmentation data are highly imbalanced. Previous studies demonstrate that neural network performance deteriorates as class imbalance increases. \unskip~\cite{406596:9001200} The resulting imbalanced data model would be highly trained for the features of the majority class but poorly trained for the features of the minority class, causing instability in early training. Despite a probable technically high accuracy from the generated model, the output results would be poor segmentations. 

A dice loss function is a type of loss function specifically designed to mitigate dataset class imbalance and is frequently used for medical imaging algorithms. \unskip~\cite{406596:8996620}  The dice score is measured as an overlap of the output mask with ground truth to assess each segmentation task and is specifically designed for use in volumetric segmentation in medical imaging. \unskip~\cite{406596:8996620} The coefficient measures the overlap between set X, the ground truth, and Y, the predicted mask. For binary class segmentation, the dice score is expressed as the following: 
\begin{eqnarray*}DiceScore=\frac{2\ast\left|X\cap Y\right|}{\left|X\right|+\left|Y\right|} \end{eqnarray*}
Intersection-over-union (IoU), also known as the Jaccard index, is a coefficient that calculates overlap in segmentation tasks. \unskip~\cite{406596:9151730} IoU also measures the overlap between set X, the ground truth, and Y, the predicted mask. Some studies indicate that IoU values typically exceed corresponding dice coefficients except when coefficient values equal 0 or 1.\unskip~\cite{406596:9151732} The IoU is calculated as the intersection between two images divided by their union, expressed as the following:

$Intersection-over-union=\frac{\left|X\cap Y\right|}{\left|X\cup Y\right|} $

If the ground truth and predicted mask are identical, then the dice coefficient and IoU both equal 1. If the ground truth and predicted mask share no elements, then both coefficients equal 0. If the ground truth and predicted mask are neither identical nor absolutely incongruent, then each coefficient value falls between 0 and 1.
    
\section{Materials and Methods}
An IRB-exempt study using 124 de-identified HIPAA-compliant lateral chest radiographs on unique patients was performed. Images were pre-processed using contrast-adaptive histogram equalization (CLAHE) to standardize the appearance and improve image  contrast. Images were subsequently down-sampled from the original size (4238 \ensuremath{\times} 3480 pixels) to 512 x 512 pixels using bilinear interpolation. Down-sampling was performed to aid back-propagation and neural network learning within graphics processor unit (GPU) memory constraints. Segmentations of visible vertebrae were manually performed  using ImageJ version 1.50i (National Institutes of Health, USA) by both a medical student (JL) and a board-certified radiologist (PL).  All segmentations of the images were verified and adjusted as needed by a board-certified radiologist (PL) 

The resulting binary mask was additionally down-sampled to 512 x 512 pixels. Ground truth-labels were color-coded with black for vertebrae and white for each image background, as seen in Figure~\ref{f-687b}. Class imbalance in the vertebral segmentations was preempted in this study with the use of a dice loss function.

\bgroup
\fixFloatSize{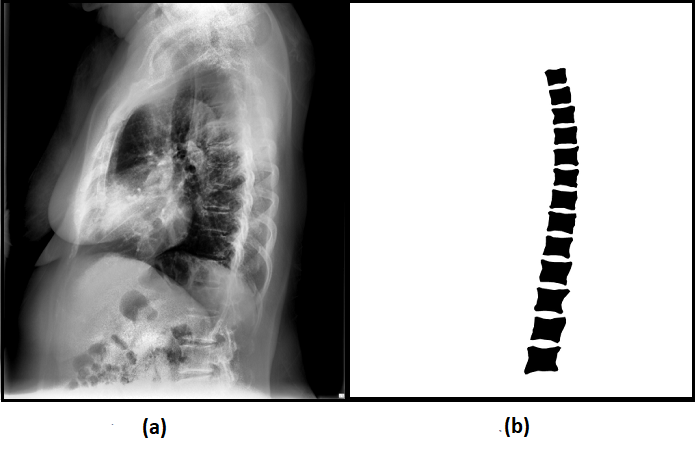}
\begin{figure*}[!htbp]
\centering \makeatletter\IfFileExists{images/0b28bc57-13cd-4779-b036-0475429f90d1-u1.png}{\includegraphics{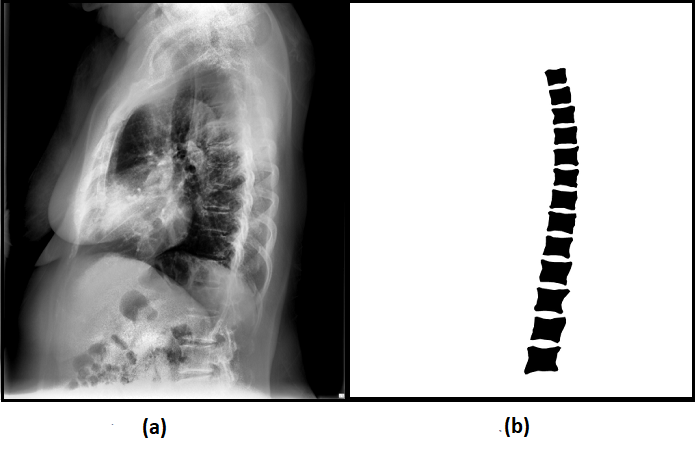}}{}
\makeatother 
\caption{{\textit{(a)\textbf{\space }Down-sampled lateral chest radiograph (b) Binary ground truth label with black vertebrae and white background}}}
\label{f-687b}
\end{figure*}
\egroup
The U-Net based convolutional neural network was employed to segment vertebrae from lateral chest radiographs. 74 images (59.68\%) were used for the training dataset, 10 images (8.1\%) were used for the validation dataset, and 40 images (32.25\%) were used for the test dataset. The model was built using Keras 2.06 \BreakURLText{(https://keras.io/)} with TensorFlow 1.1 (Google LLC, Mountain View, CA) and CUDA 8.1 (Nvidia Corporation, Santa Clara, CA). 

Differential learning rates and optimizers were trialed to enhance update weight and bias values to maximize model performance. Based on multiple experiments, the best performance resulted from using the Adam optimizer with a learning rate of 0.0001. The dice score and IoU were utilized on the validation dataset to augment model selection. The model was trained until a plateau in validation loss, which occurred at 10 epochs on a CUDA-enabled Nvidia 1080Ti 11GB graphics processing unit. (Nvidia Corporation, Santa Clara, CA) 

The dice score and IoU were used to assess model performance. As the segmentation task was binary, loss function was applied as the summation of dice score loss and binary cross entropy.
    
\section{Results}
In the holdout test dataset, an average dice coefficient value of 90.5 and an average IoU of 81.75 were obtained. Ground truth masks, predicted masks, and overlay segmentation masks were successfully generated. In Figure~\ref{f-786d}, the ground truth mask represents the manual segmentation and the predicted mask is the resulting output of the deep learning model.

\bgroup
\fixFloatSize{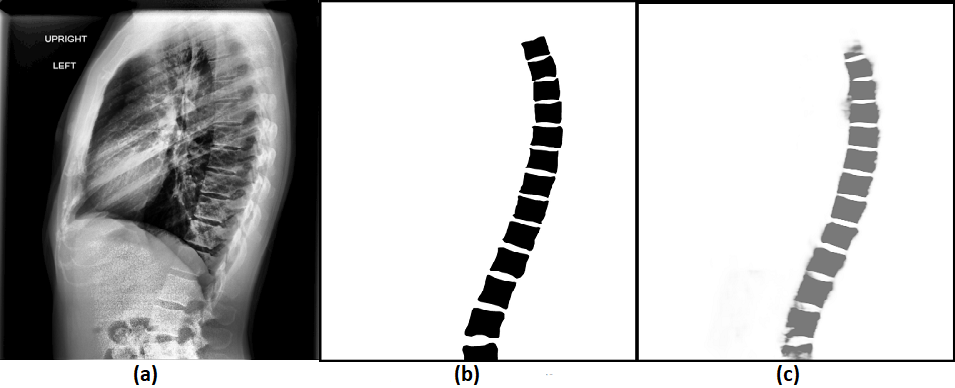}
\begin{figure*}[!htbp]
\centering \makeatletter\IfFileExists{images/811bb584-415b-48b1-a7ba-f83a60a43d83-u3-copy.png}{\includegraphics{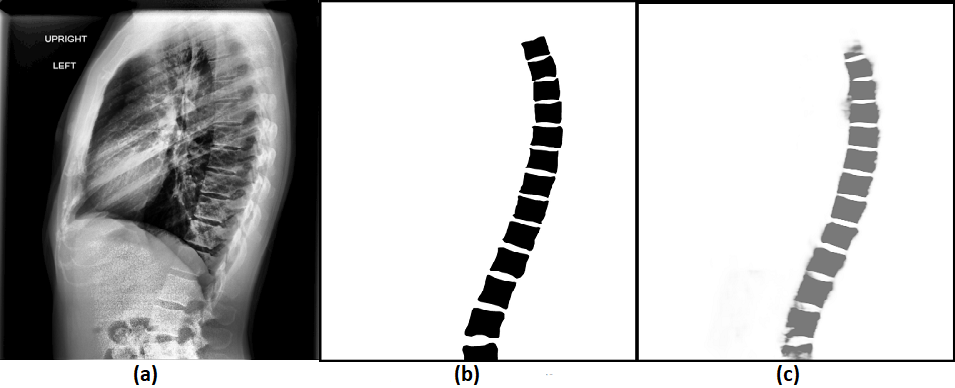}}{}
\makeatother 
\caption{{\textit{(a) Original image (b) Ground truth mask (c) Predicted mask }}}
\label{f-786d}
\end{figure*}
\egroup
In the overlay-segmented mask seen in Figure~\ref{f-b4cd}, the original radiograph is superimposed with the ground truth mask in red and the predicted mask in blue.

\bgroup
\fixFloatSize{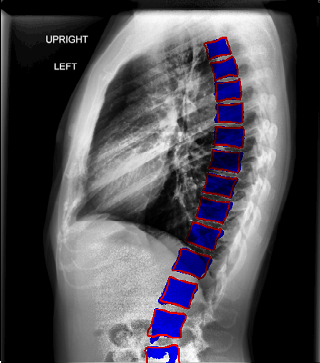}
\begin{figure*}[!htbp]
\centering \makeatletter\IfFileExists{images/1fa62215-fa6f-4175-902b-b1d996410133-u3-copy-2.png}{\includegraphics[width=.44\linewidth]{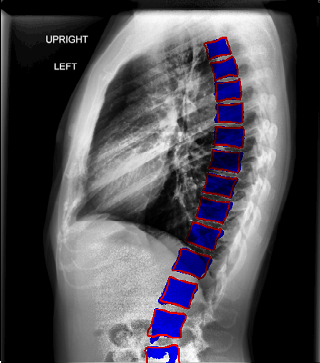}}{}
\makeatother 
\caption{{\textit{Overlay segmented mask}}}
\label{f-b4cd}
\end{figure*}
\egroup
Figure~\ref{f-24af} demonstrates an example of an inferior segmentation result. Two adjacent thoracic vertebrae, T3 and T4, as indicated by the blue arrows, are poorly visualized on the original radiograph and poorly segmented by the U-Net model. \textbf{\space }
\bgroup
\fixFloatSize{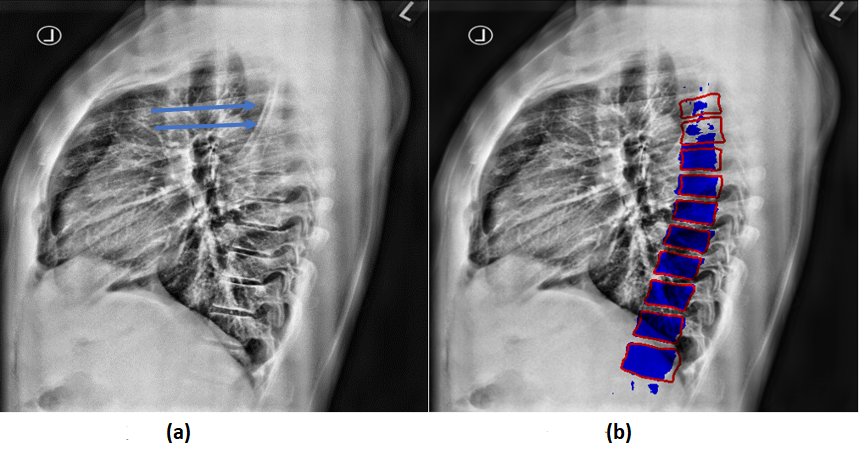}
\begin{figure*}[!htbp]
\centering \makeatletter\IfFileExists{images/253278aa-49d0-42d7-b78f-c2cad8683c41-u4.png}{\includegraphics{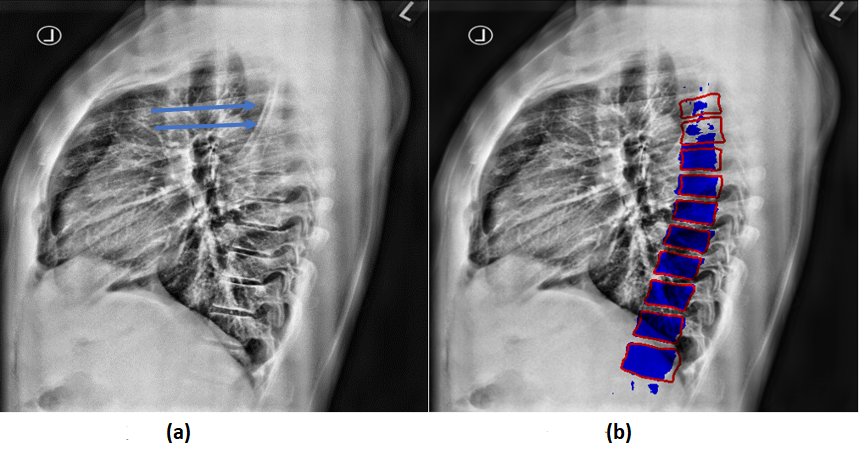}}{}
\makeatother 
\caption{{(a) \textit{Original Image (b) Overlay segmented mask }}}
\label{f-24af}
\end{figure*}
\egroup

\section{Discussion }
The U-NET segmentation network resulted in a model with a dice score of 90.12 and an IoU of 82.10 for automated segmentation of vertebrae on lateral chest radiography. These findings demonstrate that deep learning techniques for pixel-wise segmentation may require a relatively small number of images to accurately segment vertebral bodies on radiography, compared to deep learning for whole-image classification, which typically needs thousand or more images per class. \unskip~\cite{406596:12514142} This solution represents novel success with regard to a vertebral segmentation algorithm on lateral chest radiographs though there has been previous success in the classification and identification of spine fractures on 3D CT. \unskip~\cite{406596:9004732} Previous failures in automated radiographic vertebral segmentation may be due to the comparatively low-resolution of radiographs when compared with CT, especially in regard to patients with osteoporosis and other conditions that limit vertebral visualization and background contrast.  

While the U-NET was relatively successful in segmenting the thoracic vertebre, there were some cases in which the segmentation failed or was less optimal.  For example,Figure~\ref{f-24af} demonstrates one lateral radiograph where there was an improper segmentation of some of the upper thoracic vertebra.  This was  most likely due to the poor contrast resolution and relatively obscurity of the upper thoracic spine on that image related surrounding structures . As there were few cases with complex low-level spinal features, it is likely that a larger, more diverse dataset will improve the model's ability to train on more complex cases and apply that learning to complex test cases. 

Because vertebrae typically represent a minority of the total pixels on a lateral radiograph, there was a high risk for class imbalance between the segmented vertebrae and background. To combat this risk of class imbalance, the dice coefficient as a loss function was employed , and data augmentation strategies and appropriate loss functions were utilized to mitigate model overfitting. Though not explicitly assessed, it is likely that the use of the dice coefficient and data augmentation strategies improved model performance. One can consider other strategies to improve performance, including the use of a U-Net variant with Inception-inspired architecture, other FCNs, or RCNNs. \unskip~\cite{406596:9001279} A larger, more clinically diverse dataset would also likely improve performance.  

In the future, we plan to use this model in conjunction with image processing techniques that count the number of uniquely segmented vertebre starting from the first readily visible vertebrae, likely from T3 or T4, and ending at the most caudal vertebrae visible on the image.  In the future, it would be interesting to try an instance segmentation solution, such as a Mask R-CNN where each vertebra is labeled and separately segmented (e.g. T3, T4, T5, etc...). This has the advantage of a one-step deep learning solution that can identify and segment each vertebra. \unskip~\cite{406596:12514912}

As two-view chest radiography represents the most frequently utilized imaging modality in the United States, an automated screening solution to detect vertebral fractures would provide tremendous benefit to patients and could mitigate systemic healthcare expenditure. \unskip~\cite{406596:9112922}-\unskip~\cite{406596:9112921} The retrospective application of this algorithm to a large database of lateral chest radiographs could also determine normative values for mean vertebral body height, stratified by pertinent patient data, such as age, sex, height, weight and other demographic and clinical parameters.  This has implications to  better individualize patient care and prevent future spinal fractures.  
    
\section{Conclusion }
Deep learning using a U-NET demonstrates promise in the automated segmentation of vertebrae on lateral chest radiographs.

\bibliographystyle{vancouver}

\bibliography{\jobname}

\end{document}